\documentclass[iop]{emulateapj}

\usepackage{natbib}
\usepackage{amsmath}

\newcommand{\about}{$\sim\!\!$~}
\newcommand{\kms}{\,km\,s$^{-1}$}

\def\lsim{\hbox{\rlap{\raise 0.425ex\hbox{$<$}}\lower 0.65ex\hbox{$\sim$}}}
\def\gsim{\hbox{\rlap{\raise 0.425ex\hbox{$>$}}\lower 0.65ex\hbox{$\sim$}}}

\def\arcsec{\hbox{$^{\prime\prime}$}}
\def\arcdeg{\mbox{$^\circ$}}

\shorttitle{SN~2008ge: Progenitor and Supernova}
\shortauthors{Foley et~al.}

\begin{document}

\def\mag{1}
\def\gem{2}
\def\cfa{3}
\def\clay{4}
\def\harv{5}
\def\dark{6}
\def\lco{7}
\def\stock{8}
\def\and{9}
\def\chile{10}

 \title{On the Progenitor and Supernova of the SN~2002cx-like Supernova 2008ge\altaffilmark{\mag,\gem}}

\author{
{Ryan~J.~Foley}\altaffilmark{\cfa,\clay},
{Armin~Rest}\altaffilmark{\harv},
{Maximilian~Stritzinger}\altaffilmark{\dark,\lco,\stock},
{Giuliano~Pignata}\altaffilmark{\and},
{Joseph~P.~Anderson}\altaffilmark{\chile}
{Mario~Hamuy}\altaffilmark{\chile},
{Nidia~I.~Morrell}\altaffilmark{\lco},
{Mark~M.~Phillips}\altaffilmark{\lco},
and
{Francisco~Salgado}\altaffilmark{\lco}
}
%\email{rfoley@cfa.harvard.edu}

\altaffiltext{\mag}{
This paper includes data gathered with the 6.5 meter Magellan
telescope at Las Campanas Observatory, Chile.}

\altaffiltext{\gem}{
Based on observations obtained at the Gemini Observatory, Cerro
Pachon, Chile (Gemini Programs GS-2008B-Q-32, GS-2008B-Q-56, and
GS-2009A-Q-17).}

\altaffiltext{\cfa}{
Harvard-Smithsonian Center for Astrophysics,
60 Garden Street, 
Cambridge, MA 02138, USA.
}
\altaffiltext{\clay}{
Clay Fellow. Electronic address rfoley@cfa.harvard.edu .
}

\altaffiltext{\harv}{
Department of Physics,
Harvard University,
17 Oxford Street,
Cambridge, MA 02138, USA.
}

\altaffiltext{\dark}{
Dark Cosmology Centre,
Niels Bohr Institute,
University of Copenhagen,
Juliane Maries Vej 30, 2100 
Copenhagen \O, Denmark.
}

\altaffiltext{\lco}{
Carnegie Observatories,
Las Campanas Observatory,
Casilla 601, La Serena, Chile.
}

\altaffiltext{\stock}{
Department of Astronomy,
The Oskar Klein Centre,
Stockholm University,
10691 Stockholm, Sweden.
}

\altaffiltext{\and}{
Departamento de Ciencias Fisicas,
Universidad Andres Bello,
Avda.\ Republica 252,
Santiago, Chile.
}

\altaffiltext{\chile}{
Departamento de Astronom\'{i}a,
Universidad de Chile,
Santiago, Chile.
}

\begin{abstract}
We present observations of supernova (SN) 2008ge, which is
spectroscopically similar to the peculiar SN~2002cx, and its
pre-explosion site that indicate that its progenitor was probably a
white dwarf.  NGC~1527, the host galaxy of SN~2008ge, is an S0 galaxy
with no evidence of star formation or massive stars.  Astrometrically
matching late-time imaging of SN~2008ge to pre-explosion {\it HST}
imaging, we constrain the luminosity of the progenitor star.  Since
SN~2008ge has no indication of hydrogen or helium in its spectrum, its
progenitor must have lost its outer layers before exploding, requiring
that it be a white dwarf, a Wolf-Rayet star, or a lower-mass star in a
binary system.  Observations of the host galaxy show no signs of
individual massive stars, star clusters, or \ion{H}{2} regions at the
SN position or anywhere else, making a Wolf-Rayet progenitor unlikely.
Late-time spectroscopy of SN~2008ge show strong [\ion{Fe}{2}] lines
with large velocity widths compared to other members of this class at
similar epochs.  These previously unseen features indicate that a
significant amount of the SN ejecta is Fe (presumably the result of
radioactive decay of $^{56}$Ni generated in the SN), further
supporting a thermonuclear explosion.  Placing the observations of
SN~2008ge in the context of observations of other objects in the class
of SN, we suggest that the progenitor was most likely a white dwarf.
\end{abstract}


\defcitealias{Foley09:08ha}{F09}

%%%%%%%%%%%%%%%%%%%%
%%  Introduction  %%
%%%%%%%%%%%%%%%%%%%%

\section{Introduction}\label{s:intro}

In the last decade, a new class of supernovae (SNe) has been
discovered.  The class, named after its first member, SN~2002cx
\citep{Li03:02cx}, has several observational similarities to
typical SNe~Ia, but also has several distinguishing properties: low
luminosity for its light-curve shape \citep[e.g.,][]{Li03:02cx}, a
lack of a second maximum in the NIR bands \citep[e.g.,][]{Li03:02cx},
low photospheric velocities \citep[e.g.,][]{Li03:02cx}, late-time
spectra dominated by narrow permitted \ion{Fe}{2}
\citep[e.g.,][]{Jha06:02cx}, and a host-galaxy morphology distribution
highly skewed to late-type galaxies (\citealt{Foley09:08ha}; hereafter
F09; \citealt{Valenti09}). Additionally, a single member of the class,
SN~2007J, displays strong \ion{He}{1} in its spectrum
\citepalias{Foley09:08ha}.

A recent member of this class, SN~2008ha, had both a very low
photospheric velocity at maximum brightness \citep[about
4000~\kms;][]{Foley10:08ha} and extremely low luminosity ($M_{V} =
-14.2$~mag at maximum brightness; \citetalias{Foley09:08ha};
\citealt{Valenti09}).  The observations of this SN are difficult to
match to conventional SN models.  \citet{Valenti09} suggested that
SN~2008ha (and possibly all SN~2002cx-like objects) had massive
progenitors, while \citetalias{Foley09:08ha} explored both
massive-star progenitor models and models that involved a white dwarf
(WD) progenitor.  The light curves and velocities of SN~2008ha can be
matched by a model of a $13 M_{\sun}$ main-sequence mass CO star
undergoing a significant amount of fallback \citep{Moriya10}; however,
stars with a main-sequence mass of $M \lesssim 25 M_{\sun}$ are not
expected to undergo fallback \citep{Fryer99}.  The maximum-light
spectrum of SN~2008ha suggested that C/O burning occurred during the
explosion, pointing further to a WD progenitor \citep{Foley10:08ha}.

The best way to determine the progenitor system for these events would
be to unambiguously detect the progenitor (or donor) star of a SN in
pre-explosion imaging.  This technique has been very successful at
detecting nearby core-collapse SN progenitors \citep[see][for a
review]{Smartt09:review}, but has rarely been attempted for SNe~Ia
\citep{Maoz08, Nelemans08}.  For this type of study, the following data
are required: a high-resolution (to separate individual stars),
relatively deep (to detect the progenitor star) pre-explosion image of
the SN site and a SN image that is deep enough to have good detections
of several stars in common with the pre-explosion image.  A relative
astrometric solution is obtained for the two images, and then the SN
position can be precisely determined on the pre-explosion image.
Potential progenitor stars can then be identified and their properties
studied, or alternatively if no stars are consistent with the SN
position, limits can be placed on the properties of the progenitor
star.  With the appropriate observations, one can place interesting
constraints on the progenitor properties of SN~2002cx-like SNe.

SN~2008ge was discovered by the CHilean Automated Supernova sEarch
CHASE \citep{Pignata09} on 2008 October 8.27 (UT dates are used
throughout this paper) at mag 12.8 \citep{Pignata08:08ge} in NGC~1527,
an S0 galaxy with a recession velocity of 1212~km~s$^{-1}$.  Using the
surface brightness fluctuation method, \citet{Tonry01} determined that
the distance modulus for NGC~1527 is $\mu = 31.28 \pm 0.22$~mag.
Using the new Cepheid zero point of \citet{Freedman01}, this
corresponds to a corrected distance modulus of $\mu = 31.12 \pm
0.22$~mag, which we will adopt as the distance modulus for SN~2008ge.

Despite discovering SN~2008ge on 2008 October 8.27,
\citet{Pignata08:08ge} also report detecting SN~2008ge on previous
images, with the earliest detection on 2008 September 8.23 (at mag
12.4) and the last non-detection on 2008 August 21.24.  Our light
curve (presented in Section~\ref{ss:phot}) shows that SN~2008ge peaked
in $V$ around 2008 September 17.

We determined from a spectrum obtained on 2008 October 15 that
SN~2008ge was similar to the peculiar SN~2002cx 23~d past maximum
brightness \citep{Stritzinger08}.  This is consistent with our
$V$-band phase of 28~d on that date.  Given the sparse pre-maximum
data, the rise time should be between 9 -- 27~d.

NGC~1527 has been twice imaged by {\it HST} before the detection of
SN~2008ge.  In this paper, we present observations of SN~2008ge and
its host galaxy, determine the properties of potential progenitor
stars at the location of SN~2008ge, and examine the implications of
those properties in the context of the progenitors for SN~2002cx-like
objects.

%%%%%%%%%%%%%%%%%%%%
%%  Observations  %%
%%%%%%%%%%%%%%%%%%%%

\section{Observations and Data Reduction}\label{s:obs}

\subsection{Identification and {\it HST} Photometry of the Progenitor Site}

NGC~1527 was observed with {\it HST}/WFPC1 on 1993 October 20 (Program
4904; PI Illingworth) and {\it HST}/WFPC2 on 1995 January 6 (Program
5446; PI Illingworth).  The WFPC1 observation was a single 300~s
exposure with the F555W (roughly $V$) filter, while the WFPC2
observation included two exposures of 80~s each with the F606W (a
broad $V$) filter.

To determine whether the progenitor of SN~2008ge is detected in the
archival {\it HST} observations, we performed differential astrometry
using optical observations of the SN.  Observations of SN~2008ge were
obtained with
LDSS3\footnote{http://www.lco.cl/telescopes-information/magellan/instruments-1/ldss-3-1/
.} with the Magellan Clay telescope on 2009 October 16.  The images
were reduced using the SMSN pipeline as described in \citet{Rest05}
and \citet{Miknaitis07}.

The astrometry was performed with the IRAF\footnote{IRAF: the Image
Reduction and Analysis Facility is distributed by the National Optical
Astronomy Observatory, which is operated by the Association of
Universities for Research in Astronomy, Inc.\ under cooperative
agreement with the National Science Foundation.} tasks \texttt{geomap}
and \texttt{geotran} using 29 objects in common with the {\it
HST}/WFPC2 images resulting in an astrometric rms of $\sigma_{\rm
LDSS3\rightarrow HST} = 45$~mas in each coordinate.  The same
procedure was performed with the {\it HST}/WFPC1 images, which despite
its smaller field of view, and therefore only 5 stars in common, we
obtained an astrometric rms of $\sigma_{\rm GB\rightarrow HST} =
20$~mas.

\begin{figure*}
\begin{center}
\epsscale{1.15}
\plotone{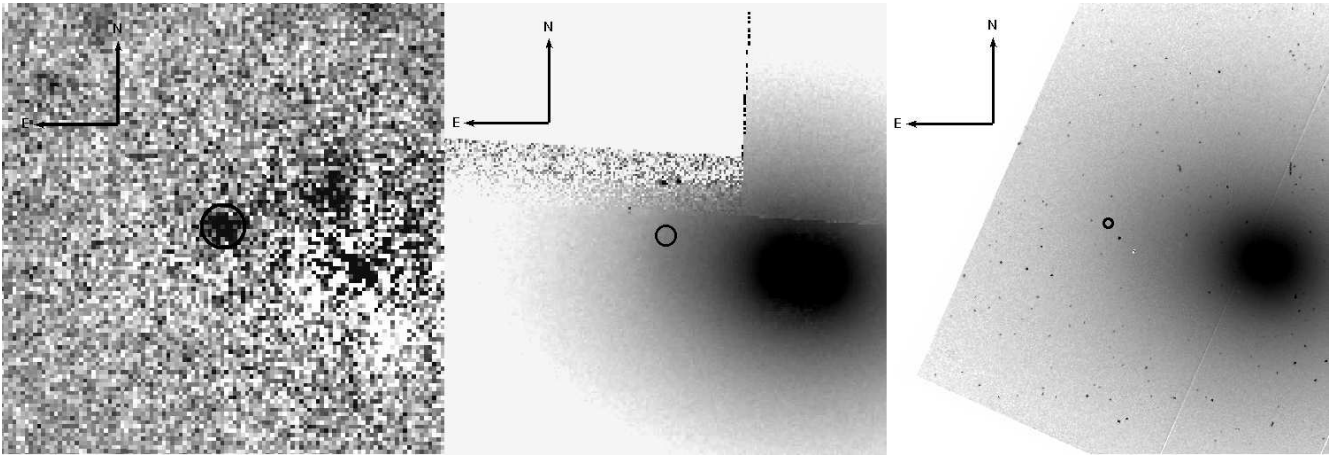}
\caption{LDSS3 $r$-band (left), {\it HST}/WFPC2 F606W (center) and
{\it HST}/WFPC F555W (right) images at the position of SN~2008ge
obtained 1 year after and 13 and 15~years before explosion,
respectively.  The images are $20\arcsec \times 20\arcsec$, and North
is up and East is left.  The LDSS3, {\it HST}/WFPC, and {\it
HST}/WFPC2 images have a pixel scale of 0.188, 0.044, and
0.1\arcsec~pixel$^{-1}$, respectively.  The LDSS3 image has been
flipped about the center of the host galaxy and subtracted from the
original image to create a difference image.  The position of
SN~2008ge is marked by the black circles whose radius corresponds to
1\arcsec for the LDSS3 image and $10 \sigma$ uncertainty in the
position for the {\it HST} images.}\label{f:proj}
\end{center}
\end{figure*}

The SN is very close to the bright nucleus of NGC~1527.  At the time
of our Magellan observation, it had faded to the point where it was
difficult to directly measure its position.  To provide an accurate SN
position, we subtracted the south-west section of the galaxy from the
north-east section of the galaxy (rotated and flipped about the galaxy
center), producing a relatively clean difference image (shown in
Figure~\ref{f:proj}).  The SN is clearly detected in our stacked
photometry with position ($\alpha$, $\delta) = ($04:08:24.689,
$-47$:53:47.01), with the absolute astrometry set using 2MASS.  This
position is consistent with that determined from our Gemini
acquisition images (when the SN was significantly brighter; see
Section~\ref{ss:spec}).  The depth of the LDSS3 images make them
preferable for determining an astrometric solution.
%4h08m24s.68 +/- 0".1, Decl. = -47o53'47".4 +/- 0".1

The position of SN~2008ge on the archival {\it HST} images are shown
in Figure~\ref{f:proj}.  In both cases, we find no clear progenitor
coincident with objects in the archival {\it HST} images.  The WFPC2
image is 0.45~mag deeper than the WFPC1 image, and the filters are
very similar.  Therefore, for the remainder of this study, we focus on
the WFPC2 image.

Since no progenitor star was detected, we can use the {\it HST}/WFPC2
image to place limits on the brightness of the progenitor.  The
surface-brightness profile is extremely smooth, and we expect that any
bright star would be easily detected.  Using \texttt{dophot}, we were
able to detect nearly all astrophysical objects to a signal-to-noise
ratio (S/N) of \about 3 with few false detections.  To determine the
exact limit, we placed artificial stars with varying brightnesses in
the image.  We then ran \texttt{dophot} on that image to recover the
artificial stars.  We find that at the SN position and positions with
similar backgrounds that we recover $>90\%$ of all objects to $S/N =
3$.  We then calculated the brightness that would yield a particular
S/N.  For $S/N = 3$ (5), we have a limiting magnitude of $m_{F606W} =
24.3$~mag (23.2~mag); at our adopted distance (and assuming no
host-galaxy extinction), this corresponds to $M_{V} = -6.7$~mag
($-7.8$~mag).

\subsection{Supernova Photometry}\label{ss:phot}

Photometry of SN~2008ge was obtained by the 0.41~m Panchromatic
Robotic Optical Monitoring and Polarimetry Telescope
\citep[PROMPT][]{Reichart05} in the Luminance filter, which is a
passband filter with wavelength range from \about 4000 to \about
7000~\AA.  Instrumental magnitudes were measured using the template
subtraction technique with a code based on the ISIS package
\citep{Alard98, Alard00}.  For calibrating the instrumental magnitudes
to the $V$ band we used the S-correction technique
\citep{Stritzinger02}, but following the prescription of
\citet{Pignata08:02dj}.  We convert to the $V$ band because its
effective wavelength is closest to the Luminance filter.  Since the
spectroscopic follow up began around 20 days after maximum light, to
correct the early photometric points we used the spectra of SN~2005hk
\citep{Phillips07}.  The two objects are both spectroscopically
similar to SN~2002cx at later times (see Figure~\ref{f:comp}) and the
correction agrees quite well for the phases where the spectroscopic
sequences overlap.  This makes us confident that the early-time
corrections are appropriate.

\begin{figure}
\begin{center}
\epsscale{1.2}
\plotone{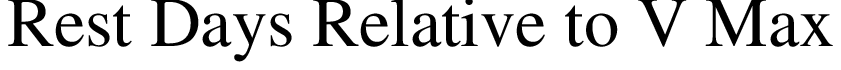}
\caption{$V$-band light curves of SNe~2002cx (dashed line), 2005hk
(solid line), and 2008ge (blue points).  The light curves of
SNe~2002cx and 2005hk have been shifted to match the light curve of
SN~2008ge at peak.}\label{f:lc}
\end{center}
\end{figure}

We present our $V$-band light curve of SN~2008ge in Table~\ref{t:phot}
and Figure~\ref{f:lc}.  The SN peaked at $V \approx 13.8$~mag on or
around 2008 September 17.  It is worth noting that SN~2005hk, which
had very well sampled light curves, peaked in $V$ 3.6~d after peaking
in $B$ \citep{Phillips07}.  At our adopted distance modulus (and
assuming no host-galaxy extinction; see Section~\ref{ss:spec}),
SN~2008ge peaked at $M_{V} \approx -17.4$~mag (with a total
uncertainty of \about 0.2--0.3~mag), comparable to
SNe~2002cx ($M_{V} = -17.49$~mag) and 2005hk ($M_{V} = -18.08$~mag;
\citealt{Phillips07}).  SN~2008ge declines slower than either
SN~2002cx or SN~2005hk in $V$.  SN~2008ge has $\Delta m_{15} (V)
\approx 0.34$~mag (the relatively large photometric uncertainty and
single data point before maximum propagates to a large uncertainty in
the measurement), which is much smaller than the measured value for
either SN~2002cx ($0.95 \pm 0.06$~mag) or 2005hk ($0.83 \pm
0.05$~mag).  We caution that our $V$ band may still be slightly
contaminated by th $R$ band, and could partially explain the width of
the $V$-band light curve.

\begin{deluxetable}{lc}
\tablewidth{0pc}
\tablecaption{CHASE $V$-Band Photometry of SN~2008ge\label{t:phot}}
\tablehead{\colhead{JD} & \colhead{$V$\tablenotemark{a} (mag)}}

\startdata

  2454717.7 & 13.95 (06) \\
  2454726.4 & 13.82 (12) \\
  2454734.8 & 13.91 (08) \\
  2454741.7 & 14.08 (06) \\
  2454747.8 & 14.40 (05) \\
  2454754.7 & 14.66 (06) \\
  2454760.6 & 14.84 (07) \\
  2454781.6 & 15.23 (05) \\
  2454787.6 & 15.34 (06) \\
  2454800.7 & 15.53 (07) \\
  2454817.6 & 15.92 (06) \\
  2454838.6 & 16.48 (08) \\
  2454842.6 & 16.45 (07) \\
  2454847.6 & 16.54 (08) 

\enddata

\tablecomments{Uncertainties (units of 0.01~mag) are given in parentheses.}

\tablenotetext{a}{$V$-band measurements are transformed from
observations observed in the Luminance filter.}

\end{deluxetable}

\subsection{Spectroscopy}\label{ss:spec}

Seven low-resolution spectra of SN~2008ge were obtained at the Las
Campanas Observatory with the Boller \& Chivens
spectrograph\footnote{http://www.lco.cl/telescopes-information/lco/telescopes-information/irenee-du-pont/instruments/website/boller-chivens-spectrograph-manuals/boller-chivens-spectrograph-manuals
.} on the 2.5~m du~Pont telescope, the IMACS spectrograph
\citep{Dressler06} on the Magellan Baade 6.5~m telescope, and the
LDSS3 spectrograph on the Magellan Clay 6.5~m telescope. An additional
three spectra were also procured with the GMOS spectrograph
\citep{Hook05} on the 8~m Gemini-South telescope.  A single
high-resolution ($R \approx 30,000$) spectrum was obtained with the
MIKE spectrograph \citep{Bernstein03} on the Magellan Clay 6.5~m
telescope.  A journal of observations is provided in
Table~\ref{t:spec}.

\begin{deluxetable}{lllrl}
%\rotate
\tabletypesize{\scriptsize}
\tablewidth{0pt}
\tablecaption{Log of Spectral Observations\label{t:spec}}
\tablehead{
\colhead{} &
\colhead{} &
\colhead{Telescope /} &
\colhead{Exposure} \\
\colhead{Phase\tablenotemark{a}} &
\colhead{UT Date} &
\colhead{Instrument} &
\colhead{(s)}}

\startdata

 40.4 & 2008 Oct.\ 27.3 & du Pont/B\&C & $3 \times 300$  \\
 46.4 & 2008 Nov.\ 2.3  & Baade/IMACS  & 300             \\
 48.3 & 2008 Nov.\ 4.2  & Clay/MIKE    & $2 \times 3600$ \\
 48.3 & 2008 Nov.\ 4.2  & Clay/LDSS3   & 300             \\
 67.4 & 2008 Nov.\ 23.3 & du Pont/B\&C & $3 \times 500$  \\
 81.2 & 2008 Dec.\ 7.1  & GS/GMOS      & $3 \times 400$  \\
114.2 & 2009 Jan.\ 9.1  & GS/GMOS      & $3 \times 600$  \\
145.2 & 2009 Feb.\ 9.1  & Clay/LDSS3   & $3 \times 600$  \\
160.2 & 2009 Feb.\ 24.1 & du Pont/B\&C & $3 \times 600$  \\
192.1 & 2009 Mar.\ 28.0 & GS/GMOS      & 2400            \\
225.1 & 2009 Apr.\ 30.0 & Clay/LDSS3   & 1800

\enddata

\tablenotetext{a}{Days since $V$ maximum, 2008 Sep.\ 16.9 (JD 2,454,726.4).}

\end{deluxetable}

Standard two-dimensional image processing and spectrum extraction was
accomplished with IRAF.  Low-order polynomial fits to calibration-lamp
spectra were derived from night-sky lines in the object frames were
applied.  Flux calibration was applied to the extracted spectra using
observations of spectrophotometric standards usually observed during
the same night as the SN.  Telluric features were removed from all of
the spectra taken at Las Campanas using high S/N spectra of telluric
standards.  Telluric corrections were not performed on the Gemini
spectra.

\begin{figure}
\begin{center}
\epsscale{1.2}
\plotone{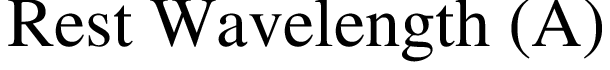}
\caption{Optical spectra of SNe~2008ge.  Phases relative to maximum
brightness are marked.}\label{f:spec}
\end{center}
\end{figure}

\begin{figure}
\begin{center}
\epsscale{1.15}
\plotone{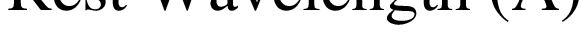}
\caption{Spectra of SN~2008ge at $+40$ (black; top panel), $+67$
(blue; bottom panel), and $+225$~d (black; bottom panel) relative to
$V$ maximum.  Comparison spectra of SN~2002cx ($+25$ and $+227$~d
relative to $B$ maximum in the top and bottom panels, respectively)
are shown in red.  For the well-sampled SN~2005hk, its light curves
peaked in $V$ 3.6~d after $B$ \citep{Phillips07}.  Line
identifications from \citet{Branch04} have been marked.}\label{f:comp}
\end{center}
\end{figure}

The spectroscopic sequence of SN~2008ge is presented in
Figure~\ref{f:spec}, and Figure~\ref{f:comp} contains a comparison
between SN~2008ge and SN~2002cx.  Although the phases in our
comparison are not perfectly matched, and some of the features appear
to be slightly broader in SN~2008ge, the strong resemblance between
the two objects is irrefutable.  The first spectrum of SN~2008ge is
dominated by Fe-group elements, with additional strong lines from Na~D
and \ion{Ca}{2}.  We refer the reader to \citet{Branch04}, which
performed a detailed study of the spectra of SN~2002cx.

No signatures of H$\alpha$ nor \ion{He}{1} is found in the spectral
series of SN~2008ge, indicating that the progenitor was likely an
evolved star.  Our high S/N, high-dispersion ($R \approx 30,000$) MIKE
spectrum of SN~2008ge shows a weak Na~D absorption system from the
Milky Way, but no evidence for Na~D absorption or H$\alpha$ emission
from the host galaxy.  We measure a $3\sigma$ upper limit of 0.01~\AA\
for the equivalent width of the Na~D2 lines suggesting that there is
little to no host-galaxy extinction.  There is also no sign of star
formation at the position of the SN nor any significant amount of
hydrogen in the circumstellar environment.

NGC~1527 is undetected in IRAS, with relatively strict limits of $<
0.1$~Jy in each band \citep{Knapp89}.  Using the \citet{Kewley02}
relationship between IRAS emission and star-formation rate (SFR), we
place a limit on the SFR of $< 7.2 \times 10^{-3}
M_{\sun}$~year$^{-1}$.  Additionally, NGC~1527 is not detected in
\ion{H}{1} 21~cm with a flux limit of $< 1.56 \times 10^{-5}$~Jy
\citep{Huchtmeier89}.  Both of these observations place significant
constraints on the SFR of the host galaxy.

%%%%%%%%%%%%%%%
%%  Results  %%
%%%%%%%%%%%%%%%

\section{Results}\label{s:results}

\subsection{Limits on Progenitor System}

\begin{figure}
\begin{center}
\epsscale{1.15}
\plotone{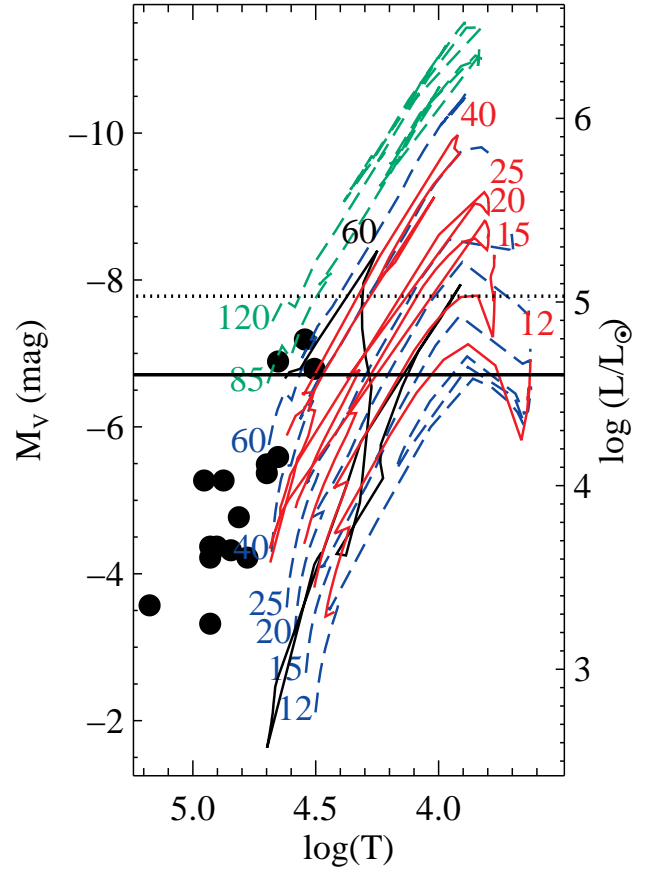}
\caption{Temperature-Magnitude diagram displaying the Geneva group
stellar evolution tracks \citep{Lejeune01} for $Z = 0.001$ (dashed
curves) and 0.1 (solid curves).  The initial mass for each track is
labeled.  The black and green curves represent initial masses that are
consistent with our magnitude limit if the star exploded on the main
sequence.  The blue and red curves represent initial masses that are
consistent with our magnitude limit if the star exploded on the main
sequence or at a later stage of evolution.  The black dots represent
Wolf-Rayet stars \citep{Crowther07} corrected from $v$ to $V$ using
the relationship of \citet{Breysacher86}.  The black solid (dotted)
horizontal line represents our $3 \sigma$ ($5 \sigma$) limiting
magnitude for the progenitor of SN~2008ge.}\label{f:cmd}
\end{center}
\end{figure}

In Figure~\ref{f:cmd}, we present the non-rotating standard mass-loss
evolutionary tracks for stars with $Z = 0.001$ and 0.1 (corresponding
to 0.05 and $5 Z/Z_{\sun}$, respectively) of the Geneva group
\citep{Lejeune01}.  We compare these tracks to the upper limit of
$M_{V} = -6.7$~mag for the progenitor of SN~2008ge.  Only knowing the
limit of the luminosity in a single band, we can place limits
on the initial mass of the progenitor and possible binary companion.
If any star in the progenitor system was still on the main sequence
and had low metallicity, then its initial mass is $M_{\rm init}
\lesssim 85 M_{\sun}$.  For high metallicity, the limit is lower with
$M_{\rm init} \lesssim 60 M_{\sun}$.  If the progenitor was on the
horizontal or red-giant branch, then there are stricter limits of
$M_{\rm init} \lesssim 12 M_{\sun}$ for both the low and high
metallicity cases.

As discussed in Section~\ref{ss:spec}, there is no indication of
hydrogen in the spectrum of SN~2008ge, so the progenitor star was
almost certainly not on the main sequence, horizontal branch, or
red-giant branch star (although a potential binary companion could
be).  In Figure~\ref{f:cmd}, we also plot example Wolf-Rayet stars
\citep{Crowther07}.  The most luminous of these stars, which
correspond to late-type WN Wolf-Rayet stars, are ruled out by our
3-$\sigma$ limit.  However, earlier WN and WC/WO stars can not be
directly ruled out by the current observations.

NGC~1527 is an S0 galaxy with no indication of star formation in its
spectrum, no indication of a dust lane or other star formation
(\citealt{Phillips96}; Figure~\ref{f:proj}), and no detection from
{\it IRAS} or in \ion{H}{1}, placing strong constraints on the SFR of
$< 7.2 \times 10^{-3} M_{\sun}$~year$^{-1}$.  Fitting a
single-stellar population \citep{Jimenez03} to the host-galaxy
spectrum (obtained at the same time as the SN spectrum), we find a
good fit with a single 9.5~Gyr population (see \citealt{Foley10:06bt}
for details of the fitting).  Additionally, there is no indication of
any emission lines in the host-galaxy spectrum, placing strong
constraints on the presence of massive stars or star clusters.  All
evidence indicates that there is not any significant population of
massive stars in NGC~1527.  The surface-brightness profile is
extremely smooth, giving further evidence of a lack of individual
massive stars or star clusters in the galaxy.  For a star brighter
than our detection limit to be masked by the galaxy, there would need
to be large fluctuations to the surface-brightness profile, which do
not exist.

\subsection{Unique Phase Spectroscopy}

Very few SN~2002cx-like objects have had their spectra published, and
only SNe~2002cx \citep{Li03:02cx}, 2005hk \citep{Phillips07, Sahu08},
2007qd \citep{McClelland10} and 2008ha (\citetalias{Foley09:08ha};
\citealt{Foley10:08ha, Valenti09}) have had their light curves
published.  As a result, only these four objects have spectra with
phase information.  \citet{Li03:02cx} published spectra with phases
ranging from $-4$ to $+56$~d relative to $B$ maximum.
\citet{Jha06:02cx} presented spectra of SN~2002cx at $+227$ and
$+277$~d relative to $B$ maximum.  \citet{Phillips07} presented
spectra of SN~2005hk that covered $-8$ to $+67$~d relative to $B$
maximum while \citet{Sahu08} extended the spectroscopic coverage by
presenting spectra at $+228$ and $+377$~d.  SN~2007qd only has four
published spectra with the earliest and latest being at $+3$ and
$+15$~d, respectively.  Although SN~2008ha had some spectroscopic
differences from SNe~2002cx and 2005hk, spectra ranging from $-1$ to
$+65$~d relative to $B$ maximum have also been published
(\citetalias{Foley09:08ha}; \citealt{Foley10:08ha, Valenti09}).
However, no spectra of this class have been published that cover the
phases 68 -- 226~d relative to $B$ maximum.

Our spectroscopic coverage of SN~2008ge fills in this gap with 6
spectra in this range.  As seen in Figure~\ref{f:spec}, during this
time the spectra remain relatively similar.  This is not unexpected
since the spectra of SN~2002cx at $+56$ and $+227$~d were very similar
to each other except for the line widths \citep{Jha06:02cx}.  The main
change in the spectral evolution of SN~2008ge is that the feature at
\about 7300~\AA\ (which we identify as [\ion{Fe}{2}]; see
Section~\ref{ss:late}) increases in strength with time.

Although it is important to see a SN~2002cx-like object transition
from early to late times, SN~2008ge may not behave in a typical manner
for this class.  Unlike SNe~2002cx and 2005hk, the widths of the
features do not significantly decrease (to separate into narrow,
primarily \ion{Fe}{2}, features) in SN~2008ge with time.

\subsection{Late-Time Spectroscopy}\label{ss:late}

At late times, the spectra of SNe~2002cx and 2005hk were dominated by
permitted \ion{Fe}{2} lines, with additional features from
\ion{Na}{1}, \ion{Ca}{2}, [\ion{Ca}{2}], and possibly [\ion{O}{1}]
\citep{Jha06:02cx, Sahu08}.  \citet{Jha06:02cx} and \citet{Sahu08}
both suggested that these objects may have emission from [\ion{Fe}{2}]
$\lambda\lambda 7155, 7453$, but alternatively considered the bluer
feature being from \ion{O}{1} $\lambda 7157$.

The 225~d spectrum of SN~2008ge is similar to SN~2002cx at $t = 227$~d
(see Figure~\ref{f:comp}), but SN~2008ge has much broader lines.
SNe~2002cx and 2005hk had extremely low-velocity features (\about
500~\kms) at these late times.  Clearly, SN~2008ge does not share this
characteristic.  Fitting a Gaussian to the relatively isolated feature
at 5900~\AA, which is likely Na~D (but could possibly be [\ion{Co}{3}]
$\lambda 5890$; \citealt{Kuchner94}), we find a FWHM of 4300~\kms.

\begin{figure}
\begin{center}
\epsscale{1.15}
\plotone{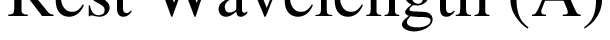}
\caption{{\it Top panel}: Late-time spectra of SNe~2002cx (red; $t =
227$~d; \citealt{Jha06:02cx}), 2005P (blue; \citealt{Jha06:02cx}), and
2008ge (black; $t = 220$~d).  The SN~2005P spectrum was obtained on
2005 May 11, but no light curve has been published for this object.
{\it Bottom panel}: Residual spectra created by subtracting the
spectra of SNe~2002cx and 2005P from SN~2008ge (red and blue curves,
respectively).  The strong residuals corresponding to [\ion{Fe}{2}]
$\lambda\lambda 7155$, 7171, 7388, 7453; [\ion{Ca}{2}]
$\lambda\lambda 7291$, 7324; and \ion{Ca}{2} NIR triplet are
marked.}\label{f:diff}
\end{center}
\end{figure}

In addition to the line widths, the main difference between the
late-time spectra of SNe~2002cx and 2008ge is the strong emission near
7300~\AA\ for SN~2008ge.  One might expect this to be [\ion{Ca}{2}]
$\lambda\lambda 7291$, 7324 since this is one of the strongest
features in the late-time spectra of SNe~2002cx \citep{Jha06:02cx},
2005hk \citep{Sahu08}, and SN~2008ha \citepalias{Foley09:08ha}.
However, subtle differences in the wavelengths of these features
require additional investigation.

By scaling the continua of the late-time spectra of SNe~2002cx and
2008ge, one can then subtract the SN~2002cx spectrum from the
SN~2008ge spectrum to obtain a residual spectrum.  This spectrum is
presented in the bottom panel of Figure~\ref{f:diff}.  As expected,
the largest residuals occur at the previously mentioned feature.
However, in the residual spectrum, it is clear that this feature is
not consistent with [\ion{Ca}{2}] $\lambda\lambda 7291$, 7324.  There
are two peaks at about 7185~\AA\ and 7395~\AA.  We interpret these
features as coming from [\ion{Fe}{2}] $\lambda\lambda 7155$, 7171,
7388, 7453.  It is also possible there is some emission from
[\ion{Ni}{2}] $\lambda 7378$.  The exact centers of these lines are
difficult to measure since the relative strengths of additional
underlying lines in the spectra of SNe~2002cx and 2008ge may be
different.  In addition to the positive residuals from [\ion{Fe}{2}],
there is a strong negative residual from the \ion{Ca}{2} NIR triplet.

Since the line widths of SNe~2002cx and 2008ge are very different at
these times, we also compare SN~2008ge to another member of the class,
SN~2005P.  No light curves of SN~2005P have been published, so the
phase information is not precise.  SN~2005P was discovered on 2005 Jan
21 by \citet{Burket05} with the last non-detection from 2004 July 8.
A spectrum obtained on 2005 May 11 and presented by \citet{Jha06:02cx}
is reproduced in Figure~\ref{f:diff}.  Assuming that SN~2005P peaked
at most 10~d after discovery, the spectrum has a phase of $100 \le t
\le 307$~d.  Although the phase of the spectrum might not be perfectly
consistent with that of SN~2008ge, it is somewhat similar (and
presumably it did not evolve quickly at these late times).

The spectra of SNe~2005P and 2008ge are very similar.  Although
SN~2005P has strong emission at the wavelengths corresponding to the
[\ion{Fe}{2}] features, they are not as strong as in SN~2008ge.
Additionally, the centroid of the red peak is offset from that of
SN~2008ge.  To further examine these differences, we produce a
residual spectrum in the same manner as above and present the result
in the bottom panel of Figure~\ref{f:diff}.  The residual spectrum has
strong positive residuals -- but not as strong as for SN~2002cx -- at
the wavelengths of [\ion{Fe}{2}], suggesting that the SN~2008ge has
stronger [\ion{Fe}{2}] emission than SN~2005P, which in turn has
stronger emission than SN~2002cx.  The residual spectrum also has a
strong negative residual at the \ion{Ca}{2} NIR triplet.

From the residual spectrum, it is clear that the offset in the
centroid of the red peak of the 7300~\AA\ feature for SN~2005P and
2008ge is the result of weaker [\ion{Fe}{2}] emission and stronger
[\ion{Ca}{2}] $\lambda\lambda 7291$, 7324 in SN~2005P.  The
[\ion{Ca}{2}] lines are easily identified as strong negative residuals
in the residual spectrum.

Nebular lines have recently been used to probe the asymmetry of SNe~Ia
\citep{Maeda10}.  In particular, the [\ion{Fe}{2}] $\lambda 7155$ and
[\ion{Ni}{2}] $\lambda 7378$ are excellent probes of the deflagration
stage of the SN explosion.  For the emission feature at 7400~\AA, it
is unclear what the relative contributions of [\ion{Ni}{2}] $\lambda
7378$, [\ion{Fe}{2}] $\lambda\lambda 7388$, 7453, and [\ion{Ca}{2}]
$\lambda\lambda 7291$, 7324 are; it is therefore difficult to assess a
velocity shift for any of these features.  However, the feature at
\about 7200~\AA\ is almost exclusively [\ion{Fe}{2}] $\lambda 7155$,
with some contribution from [\ion{Fe}{2}] $\lambda 7171$.  Fitting a
Gaussian to this line in both the normal and SN~2002cx-subtracted
spectra, we find velocity offsets (assuming that the line is dominated
by [\ion{Fe}{2}] $\lambda 7155$) of $+900$ and $+1350$~\kms,
respectively.

\citet{Maeda10} found offsets between about $-3000$ and $+3000$~\kms.
SNe~2002cx and 2005hk both had velocity offsets (as measured by the
[\ion{Ca}{2}] lines) of \about $+300$~\kms\ \citep{Sahu08}.  SN~2005hk
had relatively low polarization at maximum light, indicating a small
asymmetry \citep{Chornock06}.  It is unclear if the shift of forbidden
lines in SN~2008ge is indicative of asymmetry, and furthermore, if
such an asymmetry translates into a different late-time spectrum than
other members of the class.

%%%%%%%%%%%%%%%%%%
%%  Discussion  %%
%%%%%%%%%%%%%%%%%%

\section{Discussion \& Conclusions}\label{s:disc}

\subsection{The Progenitor of SN~2008ge}

The progenitors of SN~2002cx-like objects have recently been a subject
of debate (\citetalias{Foley09:08ha}; \citealt{Foley10:08ha,
Valenti09}).  Fortunately, the position of SN~2008ge, which we have
shown to be a spectroscopic member of the class, was imaged by {\it
HST} before the star exploded.  Since SN~2008ge is a very nearby
object, these images constrain the luminosity, and therefore mass, of
the progenitor star and any possible binary companion.

We have pinpointed the location of the SN in pre-explosion images that
indicate that the progenitor star (or system) had $M_{V} \ge -6.7$~mag
($3 \sigma$ limit), corresponding to a mass limit of $M_{\rm init}
\lesssim 85 M_{\sun}$ for a star on the main sequence and $M_{\rm
init} \lesssim 12 M_{\sun}$ for horizontal or red-giant branch star.
The same limits apply to any potential binary companion.

Since there is no indication of H or He in the spectrum of SN~2008ge,
its progenitor was likely a highly evolved star such as a WD or
Wolf-Rayet star.  Our limits are only able to rule out the
most-luminous Wolf-Rayet stars, corresponding to stars with initial
masses of $\gtrsim 65 M_{\sun}$ \citep[][and references
therein]{Crowther07}, while the minimum initial mass to reach a
Wolf-Rayet stage appears to be \about $25 M_{\sun}$.

Observations of the host galaxy also present indirect constraints on
the progenitor.  The spectrum of the galaxy has no emission lines, and
is well fit by a single-stellar population model with an age of
9.5~Gyr.  The surface-brightness profile of the galaxy is extremely
smooth, indicating that there are no exceptionally luminous stars or
large star clusters near the position of SN~2008ge.  Limits from {\it
IRAS} place the SFR below $7.2 \times 10^{-3} M_{\sun}$~year$^{-1}$.
Wolf-Rayet stars are very young, and they are generally associated
with high SFRs and spatially coincident with star clusters and
\ion{H}{2} regions \citep[e.g.,][]{Hadfield05} that should have been
detected in the {\it HST} images if present.  We see no indication of
(1) narrow emission lines in the SN spectrum, (2) narrow emission
lines in the host-galaxy spectrum, (3) far-infrared emission from the
host galaxy, (4) \ion{H}{1} 21~cm emission from the host galaxy, (5)
any luminous source at the position of the SN, or (6) any luminous
source within the {\it HST} field of view.  This evidence greatly
constrains the environment of the progenitor of SN~2008ge and is
highly suggestive that it was not a massive star of any sort.

The pre-explosion imaging combined with a lack of hydrogen in the SN
spectrum rules out all stars except for some Wolf-Rayet stars, WDs,
and relatively low-mass binary stars.  The additional observations of
the host-galaxy make a Wolf-Rayet progenitor unlikely.  A binary
system where one star has transferred its hydrogen (and possibly
helium) envelope to its companion before exploding (possibly through
electron capture) is possible.  A WD progenitor is also consistent
with all observations.  But if the progenitor of SN~2008ge was a WD,
then we expect any binary companion to have a mass less than the
maximum mass that still allows WD formation (otherwise the progenitor
of SN~2008ge would have exploded as a SN before reaching the WD
stage).  From open clusters, we see that some stars with $M_{\rm init}
= 6.5 M_{\sun}$ will become WDs \citep{Ferrario05}.  For all stages of
stellar evolution, a $6.5 M_{\sun}$ star would be below our detection
limit; therefore, this is also consistent with our observations.

\subsection{SN~2008ge: The Supernova}

Unfortunately, SN~2008ge was detected long after maximum brightness,
precluding detailed early-time observations.  Luckily, we were able to
recover the SN on our pre-detection images.  With these images, we
constructed a light curve that included maximum light.

SN~2008ge peaked at $M_{V} \approx -17.4$~mag, similar to the peak
absolute magnitude of SNe~2002cx and 2005hk, but fainter than typical
SNe~Ia.  Its $V$-band light curve declines very slowly, with $\Delta
m_{15} (V) \approx 0.34$~mag.  The similar peak magnitudes and ejecta
velocity of SNe~2002cx and 2008ge and drastically different decline
rates can be explained if both objects generated a similar amount of
$^{56}$Ni, but SN~2008ge had more massive ejecta.  More detailed
modeling, which is beyond the scope of this paper, is required to
determine the exact physical parameters of these objects.

The late-time spectra of SNe provide a glimpse at the inner regions of
the SN ejecta.  Although most SNe become ``nebular'' by 200~d after
maximum, this is clearly not the case for SNe~2002cx and 2005hk, which
had P-Cygni lines at this time, indicating a photosphere
\citep{Jha06:02cx, Sahu08}.  The late-time spectrum of SN~2008ge does
not clearly show any P-Cygni lines.  However, the overall shapes of
the spectra of SNe~2002cx and 2008ge are very similar, and it is
likely that the composition of their ejecta are very similar, but
SN~2008ge simply has a larger velocity.

The nebular spectra of SNe~Ia are dominated by forbidden Fe
transitions while SNe~Ic are dominated by \ion{Mg}{1}], [\ion{O}{1}],
[\ion{Ca}{2}], and \ion{Ca}{2} features.  The detection of strong
[\ion{Fe}{2}] emission at late times is a further indication that
SN~2008ge generated a significant amount of $^{56}$Ni (which
eventually decayed to Fe).

There is a significant offset in one of the forbidden Fe lines that
can not be explained by simple galactic motion.  This offset may be
indicative of an asymmetric explosion (at least in the core) that
could perhaps explain the spectral differences between SN~2008ge and
other members of this class at late times.

\subsection{SN~2008ge in the Context of the Class of SN~2002cx-like Objects}

\citet{Valenti09} suggested that SN~2002cx-like objects, and
particularly SN~2008ha, have massive progenitors.
\citetalias{Foley09:08ha} also presented massive progenitors as one of
many possibilities for SN~2002cx-like objects in general, and
SN~2008ha in particular (although \citealt{Foley10:08ha} showed that
SN~2008ha likely had a WD progenitor).  Since it was unlikely that
SN~2008ge had a Wolf-Rayet progenitor and since fallback SNe require a
star with $M_{\rm init} \gtrsim 25 M_{\sun}$ \citep{Fryer99}, we can
rule out the fallback scenario.  The electron capture of a star in a
binary system that has lost its outer envelopes to a binary companion
is still viable for SN~2008ge; however, other observations make this
an unlikely path for the entire class of SN~2002cx-like objects
\citepalias{Foley09:08ha}.  Of all models presented by
\citetalias{Foley09:08ha}, only one is consistent with all
observations: a deflagration of a WD, where some SN~2002cx-like events
are possibly a full deflagration of a Chandrasekhar mass WD, some are
possibly full deflagrations of a sub-Chandrasekhar mass WD, and some
are a partial deflagration of a WD that does not fully disrupt the
star.  Recent numerical models of sub-Chandrasekhar WD detonations
have been successful at producing light curves and spectra somewhat
similar to normal SNe~Ia \citep{Fink10, Pakmor10, Sim10,
vanKerkwijk10}.  These models only explore progenitors with relatively
large total mass, and additional models at lower total mass may
reproduce the features of this class.  (In fact, as the total mass
decreases, the models predict that the SN will fall off of the
Phillips relationship similar to SN~2002cx-like objects;
\citealt{Sim10}.)  Whatever is the correct model, it must explain the
strong [\ion{Fe}{2}] lines in the late-time spectra of SN~2008ge,
which suggests that a significant fraction of its ejecta is Fe, likely
the result of radioactive decay of $^{56}$Ni, although a significant
portion of the Fe may come from $^{54}$Fe or even possibly directly
synthesized $^{56}$Fe.

The detection of He (but not H) in SN~2007J \citepalias{Foley09:08ha}
is a further constraint for the progenitors of these objects,
indicating that a significant amount of He is present in at least some
of the progenitor systems.  The He may come from the WD itself or from
a binary companion, requiring that the companion be a He star.

The host-galaxy morphology distribution of this class of objects is
heavily skewed to late-type galaxies, with only SN~2008ge hosted by an
early-type galaxy \citepalias{Foley09:08ha}.  Although this
distribution is consistent with SN~1991T-like SNe~Ia
\citepalias{Foley09:08ha}, which must have WD progenitors, it is an
indication that a significant number of progenitor systems must be
relatively young.

With the constraint on the progenitor of SN~2008ge, we have a crucial
limit on the progenitors of the SN~2002cx subclass of SNe.  Future
observations of nearby SNe with deep pre-explosion imaging may further
constrain the progenitors of these objects.

\begin{acknowledgments} 

{\it Facilities:} 
\facility{Du Pont (B\&C), Gemini:South (GMOS), HST (WFPC1, WFPC2),
Magellan:Baade (IMACS), Magellan:Clay (LDSS3, MIKE), PROMPT}

\bigskip
R.J.F.\ is supported by a Clay Fellowship.  G.P.\ acknowledges support
by the Proyecto FONDECYT 11090421 and from Comit\'{e} Mixto
ESO-Gobierno de Chile.  G.P.\ and M.H.\ acknowledge support from the
Millennium Center for Supernova Science through grant P06-045-F funded
by ``Programa Bicentenario de Ciencia y Tecnolog\'{i}a de CONICYT'',
``Programa Iniciativa Cient\'{i}fica Milenio de MIDEPLAN'' and partial
support from Centro de Astrof\'{i}sica FONDAP 15010003 and by Fondecyt
through grant 1060808 from the Center of Excellence in Astrophysics
and Associated Technologies (PFB 06).

We are indebted to the staffs at the Las Campanas and Gemini
Observatories for their dedicated services.  We appreciate
conversations with E.\ Berger, R.\ Chornock, W.\ High, and B.\ Stalder
about this object.  Insightful comments from D.\ Branch were very
helpful.

This material is based upon work supported by the National Science
Foundation (NSF) under grant AST--0306969.  The Dark Cosmology Centre
is funded by the Danish NSF.

Based on observations made with the NASA/ESA Hubble Space Telescope,
and obtained from the Hubble Legacy Archive, which is a collaboration
between the Space Telescope Science Institute (STScI/NASA), the Space
Telescope European Coordinating Facility (ST-ECF/ESA) and the Canadian
Astronomy Data Centre (CADC/NRC/CSA).  Based in part on observations
obtained at the Gemini Observatory, which is operated by the
Association of Universities for Research in Astronomy, Inc., under a
cooperative agreement with the US National Science Foundation on
behalf of the Gemini partnership: the NSF (United States), the Science
and Technology Facilities Council (United Kingdom), the National
Research Council (Canada), CONICYT (Chile), the Australian Research
Council (Australia), Minist\'{e}rio da Ci\^{e}ncia e Tecnologia
(Brazil) and Ministerio de Ciencia, Tecnolog\'{i}a e Innovaci\'{o}n
Productiva (Argentina).  This research has made use of the NASA/IPAC
Extragalactic Database (NED) which is operated by the Jet Propulsion
Laboratory, California Institute of Technology, under contract with
the National Aeronautics and Space Administration.

\end{acknowledgments}

\bibliographystyle{fapj}
\bibliography{astro_refs}

%\eject

\end{document}